\documentclass[journal,twoside,web]{ieeecolor}
\usepackage{generic}
\usepackage{cite}
\usepackage{graphicx}
\usepackage{hyperref}
\hypersetup{hidelinks=true}
\usepackage{textcomp}
\usepackage{amsmath,amssymb,amsfonts}
\usepackage{algorithmic,algorithm2e,float}
\SetKwInput{KwInput}{Input}                % Set the Input
\SetKwInput{KwOutput}{Output}              % set the Output
\SetKwInput{KwRepeat}{Repeat}
\SetKwInput{KwUntil}{Until}
\SetKwInput{KwReturn}{Return}
\usepackage{amsfonts}
\usepackage{graphicx}
\usepackage{textcomp}
\usepackage{tabularx}
\usepackage{array}
\usepackage{booktabs}
\usepackage{threeparttable}  
\usepackage{multirow}
\usepackage{bm}
\usepackage{amsmath}
\usepackage{subfigure}
\usepackage{amssymb}% http://ctan.org/pkg/amssymb
\usepackage{pifont}% http://ctan.org/pkg/pifont
\usepackage{adjustbox} %

\def\BibTeX{{\rm B\kern-.05em{\sc i\kern-.025em b}\kern-.08em
    T\kern-.1667em\lower.7ex\hbox{E}\kern-.125emX}}
\begin{document}
\title{EEG-Deformer: A Dense Convolutional Transformer for Brain-computer Interfaces}
\author{Yi~Ding,~\IEEEmembership{Member,~IEEE,}
Yong~Li, Hao~Sun, Rui~Liu,
Chengxuan~Tong,~\IEEEmembership{Graduate Student Member,~IEEE,}
Chenyu~Liu, Xinliang~Zhou,
and~Cuntai~Guan,~\IEEEmembership{Fellow,~IEEE}% <-this % stops a space
\thanks{Yi Ding and Yong Li contribute equally to this work.}
\thanks{Yi Ding, Yong Li, Rui Liu, Chengxuan Tong, Chenyu Liu, Xinliang Zhou, and Cuntai Guan are with the College of Computing and Data Science, Nanyang Technological University, 50 Nanyang Avenue, Singapore, 639798.}
\thanks{Hao Sun is with the Key Laboratory of Smart Manufacturing in Energy
Chemical Process, Ministry of Education, East China University of Science and
Technology, Shanghai, China.}
\thanks{Chengxuan Tong is with Wilmar International, Singapore.}
\thanks{Cuntai Guan is the Corresponding Author.}
\thanks{This work was supported by the RIE2020 AME Programmatic Fund, Singapore (No. A20G8b0102) and the MoE AcRF Tier 1  Project (No. RT01/21)}
}
\maketitle
\begin{abstract}
Effectively learning the temporal dynamics in electroencephalogram (EEG) signals is challenging yet essential for decoding brain activities using brain-computer interfaces (BCIs). Although Transformers are popular for their long-term sequential learning ability in the BCI field, most methods combining Transformers with convolutional neural networks (CNNs) fail to capture the coarse-to-fine temporal dynamics of EEG signals. To overcome this limitation, we introduce EEG-Deformer, which incorporates two main novel components into a CNN-Transformer: (1) a Hierarchical Coarse-to-Fine Transformer (HCT) block that integrates a Fine-grained Temporal Learning (FTL) branch into Transformers, effectively discerning coarse-to-fine temporal patterns; and (2) a Dense Information Purification (DIP) module, which utilizes multi-level, purified temporal information to enhance decoding accuracy. Comprehensive experiments on three representative cognitive tasks—cognitive attention, driving fatigue, and mental workload detection—consistently confirm the generalizability of our proposed EEG-Deformer, demonstrating that it either outperforms or performs comparably to existing state-of-the-art methods. Visualization results show that EEG-Deformer learns from neurophysiologically meaningful brain regions for the corresponding cognitive tasks. The source code can be found at https://github.com/yi-ding-cs/EEG-Deformer.
\end{abstract}


\section{Introduction}
\IEEEPARstart{B}{rain}-computer interface (BCI) technology facilitates direct communication between the brain and machines using electroencephalography (EEG) \cite{9093122}. A standard BCI system usually consists of four key components: data acquisition, pre-processing, classification, and feedback \cite{lotte2010regularizing}. BCIs are employed in various practical applications, such as stroke rehabilitation \cite{foong2019assessment}, sleep stage detection\cite{10663937,10.1145/3637528.3671981}, and emotion regulation in mental health treatments \cite{zotev2020emotion}. 

\begin{figure}[t]
    \centering
    \includegraphics[width= \linewidth]{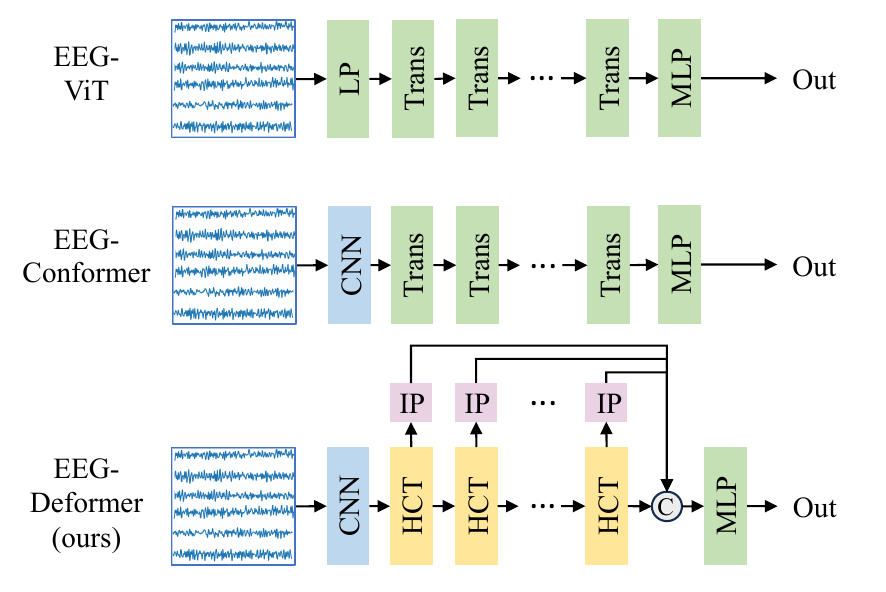}
    \caption {Comparison of network architectures between ViT \cite{dosovitskiy2021an}, EEG-Conformer \cite{9991178}, and our proposed \textbf{EEG-Deformer}. We propose a novel Hierarchical Coarse-to-Fine Transformer (HCT). Additionally, we have designed Information Purification Unit (IP-Unit, denoted by IP in the figure) for each HCT layer with dense connections to further boost EEG decoding performance.}
    \label{fig:networks_compare}
\end{figure}

EEG signals, collected through electrodes placed on each subject's head, comprise spatial and temporal dimensions. The spatial dimension relates to the locations of the EEG electrodes, while the temporal dimension captures fluctuations in brain activity \cite{lotte2010regularizing}. To ensure accuracy in decoding neural activity, a reliable BCI system must effectively perceive the subtle temporal dynamics encoded in EEG signals, which implicitly represent various cognitive processes. However, this task is quite challenging, as these brain activities vary across subjects and are susceptible to external interference, such as movement artifacts, eye blinks, and environmental factors \cite{9658165}.

Recently, numerous deep learning-based approaches have been developed for decoding brain activities from EEG signals. These approaches can be broadly categorized into two types: methods using hand-crafted features and methods using EEG directly as input. The former \cite{8320798,9361688,10340644,10506986,10.1145/3394171.3413724} involve extracting various types of features from EEG signals for neural network input. Meanwhile, CNN-based methods, which leverage automatic feature learning, typically process EEG as 2-D time series \cite{doi:10.1002/hbm.23730,Lawhern_2018,9762054}. Given that EEG data is inherently graph-structured, with nodes representing electrodes and connections based on spatial distance, functional connectivity, or learned relationships, graph-based methods have gained popularity \cite{8320798,10025569,10.1145/3474085.3475583}. However, using features as input may result in the loss of fine-grained temporal information by averaging data along the temporal dimensions. Additionally, CNN-based methods may fail to capture long-term temporal dependencies by employing CNN kernel along temporal dimension.

In addition to CNNs, transformer-based neural architectures have attracted significant attention in the BCI field due to their inherent ability to perceive global dependencies. Commonly, prior works \cite{9845479,CHEN2023521,9735124,9991178} adopt a CNN-Transformer architecture, where the CNN part serves as an adaptive feature encoder to preprocess EEG data, and the subsequent transformer part captures long-range temporal characteristics. The CNN-based shallow feature encoder has been verified as essential for preparing EEG data for Transformers \cite{9845479}.
Although previous works can capture either fine-grained (short-period) or coarse-grained (global) temporal dependencies within each layer, they have not explicitly captured both coarse- and fine-grained temporal dynamics within the Transformer layers, which may limit the full utilization of EEG signals' long-short period temporal dynamics \cite{https://doi.org/10.1002/hbm.24949}. Moreover, most existing methods overlook the abundant latent features encoded within intermediate neural layers. These layers encapsulate rich temporal information that can be systematically explored to enhance the precise discernment of temporal dynamics inherent in EEG data.

To mitigate the above-mentioned issues and enhance the perception of temporal dynamics in EEG data, we introduce EEG-Deformer, a novel dense convolutional Transformer. We propose a novel Hierarchical Coarse-to-Fine Transformer (HCT) block that integrates a Fine-grained Temporal Learning (FTL) branch into Transformers. Built upon a CNN-based shallow feature encoder comprising collaborative temporal and spatial convolutional layers, HCT concurrently captures coarse- and fine-grained temporal dynamics, as shown in Figure~\ref{fig:networks_compare}. The FTL branch employs a 1-D CNN to sequentially capture short-period temporal dynamics of EEG, generating fine-grained temporal representations. These representations are then adaptively fused with the coarse-grained temporal representations encoded by the Transformers, thereby providing more discriminative long- and short-term temporal information.

Furthermore, to efficiently utilize multi-level temporal information from intermediate HCT layers, we have designed a Dense Information Purification (DIP) module in our EEG-Deformer. This module enables the dense transmission of multi-level representations from HCT layers to the final representation. Differing from the skip connections in DenseNet \cite{8099726}, we introduce an Information Purification Unit (IP-Unit) that transforms the fine-grained representations from FTL branches using a logarithmic (log) power operation \cite{10025569} and elegantly fuses these latent representations into the final representation. Because log power can represent the amount of activity in filtered signals and reduce dimensions \cite{nunez2006electric}, our IP-Unit not only retains critical frequency characteristics related to brain activity but also effectively reduces the number of learnable parameters. As illustrated in Figure~\ref{fig:networks_compare}, a series of IP-Units have been added to the intermediate layers. These units progressively encode discriminative temporal representations. We will validate these enhancements in Sec.~\ref{sct:ablation} to \ref{sct:location_IP}.

In summary, the contributions of this work are summarized as follows:
\begin{itemize}
\item We introduce EEG-Deformer, a novel architecture designed for EEG decoding across various cognitive tasks.
\item A hierarchical coarse-to-fine Transformer is proposed to effectively encode the coarse-to-fine temporal dynamics within EEG data.

\item We develop a dense information purification module to fully exploit abundant intermediate multi-level EEG features and enhance the EEG decoding performance.

\item Through extensive experimentation on three public datasets, encompassing attention, fatigue, and cognitive workload classification tasks, the efficacy of EEG-Deformer is demonstrated. Our results show its superiority compared to current state-of-the-art (SOTA) methods. 
\end{itemize}

\begin{figure*}[t]
    \centering
    \includegraphics[width= \linewidth]{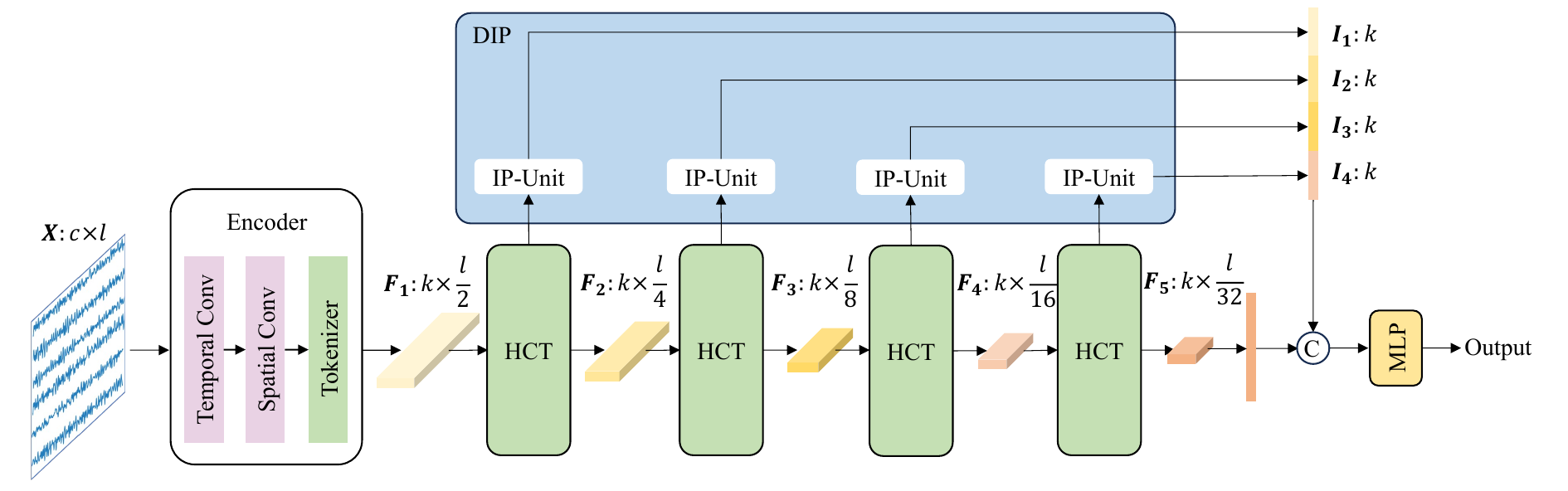}
    \caption{The network structure of EEG-Deformer. EEG-Deformer consists of three main parts: (1) Shallow feature encoder, (2) Hierarchical coarse-to-fine-Transformer (HCT), and (3) Dense information purification (DIP). The fine-grained representations from each HCT will be passed to Information Purification Unit (IP-Unit) and concatenated (C) to the final embedding.}
    \label{fig:Deformer}
\end{figure*}

\section{Related Work}
\subsection{CNNs for EEG Decoding}
Convolutional Neural Networks (CNNs) have become powerful tools in BCI applications, excelling in learning directly from EEG data \cite{doi:10.1002/hbm.23730,Lawhern_2018,9762054,WANG2024312}. Schirrmeister et al. \cite{doi:10.1002/hbm.23730} introduced DeepConvNet, which incorporates a two-stage spatial and temporal convolution layer to facilitate EEG feature extraction and classification. Similarly, Lawhern et al. \cite{Lawhern_2018} developed EEGNet, using depth-wise convolution with a kernel size of $(n,1)$—where $n$ represents the number of EEG channels—to capture spatial features. Building on these methods, TSception \cite{9762054} employs multi-scale convolutional kernels to decode temporal dynamics and asymmetric spatial activations within EEG signals. However, the relatively short length of these 1-D CNN kernels in the temporal dimension limits their ability to capture long-term temporal patterns effectively.

\subsection{Transformers for BCI}

Transformers, renowned for capturing long-term dependencies in sequential data, have gained considerable attention in EEG research \cite{9845479,CHEN2023521,9735124,9991178,CHENG2024106624,10261214}. Lee et al. \cite{9735124} showed that integrating a self-attention module from the Transformer architecture into an EEGNet-based CNN improves classification accuracy for imagined speech from EEG data. EEG-Conformer \cite{9991178}, a streamlined convolutional Transformer model, effectively combines local and global features for EEG decoding within a unified structure. However, most existing convolutional Transformers, which typically leverage CNNs for shallow feature extraction followed by Transformer blocks, may struggle to learn both coarse- and fine-grained temporal patterns in EEG signals. Furthermore, they frequently miss out on capturing multi-level temporal information available across different layers.

\section{Method}
In this work, we propose a novel Transformer, EEG-Deformer, for general EEG decoding in the BCI field. The network architecture is shown in Figure ~\ref{fig:Deformer}. EEG-Deformer consists of three main components: (1) a Shallow feature encoder, (2) a Hierarchical Coarse-to-fine Transformer (HCT), and (3) a Dense Information Purification (DIP) module. Given an input segment of EEG signals, EEG-Deformer utilizes the CNN feature encoder to adaptively encode the shallow temporal and spatial features, which are then set as input into the following HCT blocks to extract the temporal dynamics that happen in different timescales in EEG signals. To effectively perceive the critical multi-level temporal information, the features generated from each HCT block are adaptively fused via progressive IP-Units. Below, we present the details for each of them.

\begin{figure}[t]
    \centering
    \includegraphics[width=0.8\linewidth]{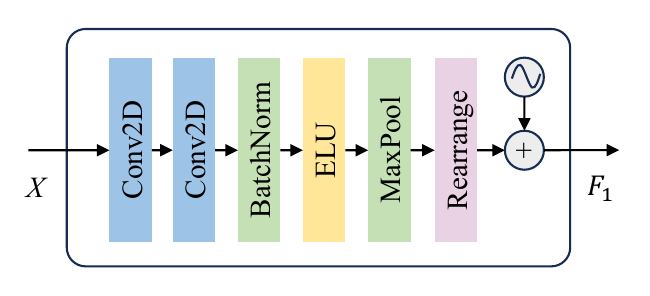}
    \caption{The structure of the shallow feature encoder. After standard CNN layers, the representation is rearranged into kernel by feature, and a position encoding is added onto it.}
    \label{fig:encoder}
\end{figure}
\subsection{Shallow Feature Encoders}
To capture the shallow temporal and spatial information of EEG, we utilize a CNN-based shallow feature encoder. It comprises two main components: CNN layers and a tokenizer. CNNs along temporal and spatial dimensions are commonly used as feature extractors for EEG signals \cite{Lawhern_2018}.
In our proposed EEG-Deformer, a two-layer CNN is adopted as the shallow feature encoder. The architecture of the CNN encoder is shown in Figure ~\ref{fig:encoder}. It begins with temporal and spatial CNN layers, followed by batch normalization to mitigate the covariate shift issue. 

Let's denote an EEG sample as $\textit{\textbf{X}} \in \mathbb{R}^{c \times l}$, where $c$ is the number of EEG channels, and $l$ represents the number of data points along the time dimension. Inspired by neurophysiological knowledge that the brain has microstates lasting approximately 100 ms, the temporal CNN kernels are sized at $(1, 0.1 \times f_{s})$,  where $f_{s}$ denotes the EEG's sampling rate. For capturing spatial information from EEG, a spatial CNN kernel of size $(c, 1)$ is utilized as suggested in \cite{Lawhern_2018}, where $c$ is the number of EEG channels. Weight normalization is included following the approach in \cite{9175874}. The number of CNN kernels is denoted by $k$. After activation by the ELU function, the learned features are max-pooled every two data points without overlapping. 
The tokenizer includes rearrange operation and a learnable position encoding. 
The size of the features is then rearranged into $k \times 0.5l$ to serve as tokens for the Transformers, which learn coarse-grained temporal information. Subsequently, a learnable position encoding, $\textit{\textbf{P}} \in \mathbb{R}^{k \times 0.5l}$, is added to the tokens \cite{dosovitskiy2021an}. Therefore, the encoded tokens can be represented as $\textit{\textbf{F}} \in \mathbb{R}^{k \times 0.5l}$. This step can be formularized as

\begin{equation}\label{eq:cnn_encoder}
      \boldsymbol{F} = \Gamma(\textrm{MaxPool}(\textrm{ELU}(\textrm{BN}(\textrm{CNN}(\boldsymbol{X}))))) + \boldsymbol{P},
\end{equation}
where $\Gamma(\cdot)$ is the rearrange operation and $\textrm{BN}(\cdot)$ is the batch normalization layer.

\subsection{Hierarchical Coarse-to-fine Transformer}
\begin{figure}[t]
    \centering
    \includegraphics[width=0.9\linewidth]{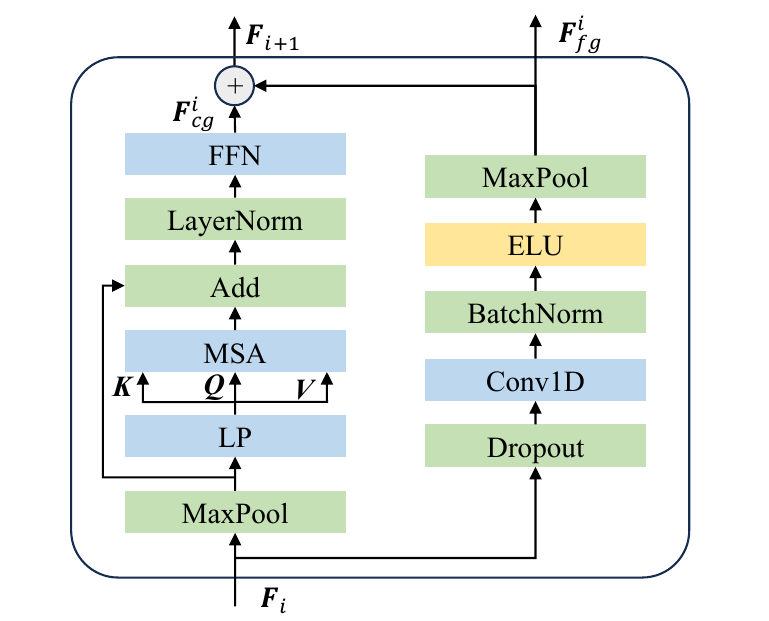}
    \caption{The structure of the hierarchical coarse-to-fine Transformer: The right side is capable of learning coarse-grained temporal information through self-attention, while the left side is the newly added FTL.}
    \label{fig:hct}
\end{figure}

With the encoded shallow EEG features, we aim to learn the coarse and fine-grained temporal dynamics in EEG data via cascading HCT blocks. The neural structure of a HCT block is shown in Fig.~\ref{fig:hct}. 
A HCT consists parallel Transformer-based branch that aims to learn the correlations among the given tokens \cite{NIPS2017_3f5ee243} and CNN-based branch that aims to learn fine-grained EEG features.

Let us suppose $\textit{\textbf{F}}_{i} \in \mathbb{R}^{k \times l_{t}^{i}}$ the input to the $i$-th HCT block. To capture the coarse-grained temporal dynamics of EEG signals, we treat the output of each CNN kernel as one token. By doing so, the long-term temporal information is explicitly included when we project the tokens into $\textit{\textbf{Q}}_{i} \in \mathbb{R}^{n_{head} \times k \times d_{attn}}$, $\textit{\textbf{K}}_{i} \in \mathbb{R}^{n_{head} \times k \times d_{attn}}$, and $\textit{\textbf{V}}_{i} \in \mathbb{R}^{n_{head} \times k \times d_{attn}}$ using the linear projection (LP) layer parameterized by $n_{head}$ set of $\textit{\textbf{W}}_{qkv}^{i} \in \mathbb{R}^{\frac{l_{t}^{i}}{2^{i-1}} \times 3d_{attn}}$. Before the LP layer, a max pooling layer with a pooling size of 2 and a step of 2 is added to reduce the feature dimension of $\textit{\textbf{F}}_{i}$. The encoding process $\Phi(\cdot)$ can be formularized by

\begin{equation}\label{eq:Emb_qkv}
      \{\boldsymbol{Q}_{i}, \boldsymbol{K}_{i}, \boldsymbol{V}_{i}\}=\Phi_{i}(\boldsymbol{F}_{i}) = \textrm{MaxPool}(\boldsymbol{F}_{i})\boldsymbol{W}^{i}_{qkv}.
\end{equation}

A multi-head self-attention (MSA) is then utilized to extract the correlations among the different views of the coarse-grained temporal embeddings. The scaled dot-product is utilized as the attention operation along temporal tokens for each attention head.

\begin{equation}\label{eq:attention}
      \textrm{Attn}(\boldsymbol{Q}, \boldsymbol{K}, \boldsymbol{V}) =\textrm{Softmax}(\boldsymbol{Q}\boldsymbol{K}^{T}/\sqrt{d})\boldsymbol{V},
\end{equation}
where $d=d_{attn}$ is a scaling factor.
The output of each attention head will be concatenated and linearly projected into a tensor, $\boldsymbol{F}_{msa}^{i}$, that has the same size of the input by

\begin{equation}\label{eq:MSA}
\begin{aligned}
    \boldsymbol{F}_{msa}^{i} = &\textrm{MSA}(\boldsymbol{F}_{i})\\
    = &[\textrm{Attn}(\Phi_{i}^{1}(\boldsymbol{F}_{i})), ..., \textrm{Attn}(\Phi_{i}^{n_{head}}(\boldsymbol{F}_{i}))]\boldsymbol{W}_{attn}^{i},
\end{aligned}
\end{equation}
where $[ \cdot]$ is concatenation operation and $\boldsymbol{W}_{attn}^{i} \in \mathbb{R}^{n_{head}d_{attn} \times \frac{l_{t}^{i}}{2^{i-1}}}$ is the learnable weights.

After the MSA, a residual connection is introduced, followed by layer normalization and a Feedforward Neural Network (FFN). The FFN consists of two layers, utilizing a GELU non-linearity, similar to those used in ViT \cite{dosovitskiy2021an}. Consequently, the coarse-grained embedding, denoted as $\boldsymbol{F}_{cg}^{i}$, is generated by:

\begin{equation}\label{eq:cg_f}
 \boldsymbol{F}_{cg}^{i} = \textrm{FFN}(\textrm{LN}(\boldsymbol{F}_{msa}^{i} + \textrm{MaxPool}(\boldsymbol{F}_{i}))),
\end{equation}
where $\textrm{LN}(\cdot)$ is the layer normalization operation. 

We further propose a FTL module to capture the short-period temporal dynamics of EEG signals. To learn fine-grained temporal features, we opt for a 1-D CNN, as its short kernel moves along the temporal dimension step by step, enabling the learning of short-period patterns. After a dropout layer, the learned representations are fed into a 1-D CNN layer, followed by a batch normalization layer, an ELU activation, and a max-pooling layer. The fine-grained temporal representations, denoted as $\boldsymbol{F}_{fg}^{i}$, can be calculated as follows:

\begin{equation}\label{eq:fg_f}
 \boldsymbol{F}_{fg}^{i} = \textrm{MaxPool}(\textrm{ELU}(\textrm{BN}(\textrm{CNN}(\textrm{DP}(\boldsymbol{F}_{i})))),
\end{equation}
where $\textrm{DP}(\cdot)$ is the dropout operation. 

After learning the coarse- and fine-grained temporal representations, a sum fusion is added to get the final output of the HCT layer

\begin{equation}\label{eq:out_hct}
 \boldsymbol{F}_{i+1} = \boldsymbol{F}_{cg}^{i} + \boldsymbol{F}_{fg}^{i}.
\end{equation}
For the $\boldsymbol{F}_{fg}^{i}$, it is also used to do the information purification process.

\subsection{Dense Information Purification}
With the learned fine-grained temporal features from different HCT layers, we further utilize densely connected IP-Units to progressively extract the multi-level temporal information from these layers. It aims to extract discriminative information from a frequency perspective and reduce the size of the bypassed multi-level representations.

Drawing inspiration from neural engineering, where power features of EEG signals in different frequency bands are widely used for brain activity analysis \cite{553713,doi:10.1002/hbm.23730}, we propose a power layer for information purification and encode frequency information in EEG signals. The power of the learned 1-D hidden representations , $\boldsymbol{I}_{i} \in \mathbb{R}^{k}$, can be calculated by
\begin{equation}\label{eq:power}
\begin{aligned}
    \boldsymbol{I}_{i} = &\textrm{IP}_{power} (\boldsymbol{F}_{fg}^{i}) \\
    = &\{\textrm{log}(\frac{1}{l_{t}^{i}}\sum (\boldsymbol{f}_{fg}^{i, j})^{2}): \boldsymbol{f}_{fg}^{i, j} \in \boldsymbol{F}_{fg}^{i}\},
\end{aligned}
\end{equation}
where $\textrm{log}(\cdot)$ is the logarithm, and $\boldsymbol{f}_{fg}^{i, j} \in \mathbb{R}^{l_{t}^i}$ is one learned representation from the $j$-th CNN kernel in the FTL.

The purified information from all HCT layers is concatenated with the flattened output of the last HCT layer to obtain the final hidden embedding. A linear layer is then utilized as the classifier to project the hidden embedding onto the class labels. Let $n_{hct}$ denote the total number of HCT layers; this process can be formulated as:

\begin{equation}\label{eq:out_final}
 \boldsymbol{out} = [\boldsymbol{F}_{n_{hct}+1}, \boldsymbol{I}_{1}, \dots, \boldsymbol{I}_{n_{hct}}]\boldsymbol{W} + \boldsymbol{b},
\end{equation}
where $[\cdot]$ is the concatenation operation, $\boldsymbol{W}$ and $\boldsymbol{b}$ are the trainable weights and biases.

\begin{table*}[htp] \centering\arraybackslash
\caption{Generalized cross-subject classification results from various methods on three benchmarking datasets. The best results are highlighted in bold and the next best are marked using underlines}
\label{tab:classification_results}
\begin{adjustbox}{center}
\begin{tabular}{lwc{2.8em}wc{2.8em}wc{2.8em}wc{2.8em}wc{2.8em}wc{2.8em}wc{2.8em}wc{2.8em}wc{2.8em}wc{2.8em}wc{2.8em}wc{2.8em}}
\toprule
\multirow{2}{*}{Method} & \multicolumn{4}{c}{Dataset I: Attention} & \multicolumn{4}{c}{Dataset II: Fatigue} & \multicolumn{4}{c}{Dataset III: Mental Workload} \\ \cmidrule(lr){2-5} \cmidrule(lr){6-9} \cmidrule(lr){10-13}
                & ACC & std & F1-macro & std& ACC &std & F1-macro &std & ACC & std & F1-macro & std\\ \midrule
		     DGCNN &60.98 &6.05	&56.81 &9.51 &69.56 &12.50 &64.92 &13.08 &64.85 &17.90 &59.42 &22.25\\
              LGGNet &67.81 &6.38	&66.70 &7.69 &68.16	&13.16 &65.38 &12.40 &65.37 &12.78 &61.91 &16.16\\
              EEGNet & 75.28 &9.77 &73.81 &12.62 &73.95 &12.54 &71.43 &13.45 &65.85 &14.55	&61.89 &18.21 \\
              TSception &71.60 &7.66 &69.94 &10.88 &73.01 &12.93 &70.18 &13.90 &\underline{69.92} &15.07 &65.26 &20.04 \\
              EEG-ViT &67.72 &7.03 &66.79	&8.49 &66.95 &14.12 &64.12 &13.89	&63.12 &15.44 &61.19 &16.56\\
              SSVEPformer &73.18&7.87&72.7&8.54&55.02&10.95&52.00&9.87&55.17&6.71&54.25&7.01\\
              EEG-Transformer&75.39&7.92&74.93&8.61&65.53&11.2&63.86&12.84&64.46&14.91&64.28&14.94\\
              EEG-Conformer &\underline{79.81} &9.27 & \underline{79.01} &10.75 & \underline{74.36}	&11.82 &\underline{71.65} &12.96	&69.40 &14.48 &\underline{65.59} &18.68\\ \midrule
              EEG-Deformer (ours) &\textbf{82.72} &8.00 &\textbf{82.36}	&8.52	&\textbf{79.32} &7.87 &\textbf{75.83}	&11.54 &\textbf{73.18} &15.63	&\textbf{69.99}	&19.80\\ \bottomrule
			\end{tabular}
		\end{adjustbox}
	\end{table*}

\begin{table*}[htp] \centering\arraybackslash
\caption{Results of the ablation study conducted on three benchmarking datasets, with the best results marked in bold.}
\label{tab:ablation_results}
\begin{adjustbox}{center}
\begin{tabular}
{lwc{2.8em}wc{2.8em}wc{2.8em}wc{2.8em}wc{2.8em}wc{2.8em}wc{2.8em}wc{2.8em}wc{2.8em}wc{2.8em}wc{2.8em}wc{2.8em}}

% {lwc{2.3em}wc{2.3em}wc{2.3em}wc{2.3em}wc{2.3em}wc{2.3em}wc{2.3em}wc{2.3em}wc{2.3em}wc{2.3em}wc{2.3em}wc{2.3em}}
\toprule
\multirow{2}{*}{Method} & \multicolumn{4}{c}{Dataset I: Attention} & \multicolumn{4}{c}{Dataset II: Fatigue} & \multicolumn{4}{c}{Dataset III: Mental Workload} \\ \cmidrule(lr){2-5} \cmidrule(lr){6-9} \cmidrule(lr){10-13}
                & ACC & std & F1-macro& std& ACC& std & F1-macro & std & ACC& std & F1-macro & std\\ \midrule
		     w/o FTL &70.39	&6.89 &69.89 &7.39 &70.48 &14.76 &68.63 &13.97 &63.31 &11.67	&61.79 &13.25\\
              w/o Dense connection  &75.84 &7.24 &75.40 &7.67	&70.85 &14.11	&68.86 &14.22 &72.46 &14.73	&69.30 &18.88\\
              w/o IP-Unit & 72.01 & 7.26 &71.28 & 8.06 & 73.28 & 12.10 & 70.98 & 12.64 & 67.29 & 11.75 & 64.48 & 14.73 \\
              EEG-Deformer (ours) &\textbf{82.72} &8.00 &\textbf{82.36}	&8.52	&\textbf{79.32} &7.87 &\textbf{75.83}	&11.54 &\textbf{73.18} &15.63	&\textbf{69.99}	&19.80\\ \bottomrule
			\end{tabular}
		\end{adjustbox}
	\end{table*}

\section{Experiment}
\subsection{Datasets and Pre-processing}
We evaluate the performance of EEG-Deformer with three benchmarking EEG datasets, which are Dataset I \cite{Shin2018} for cognitive attention, Dataset II \cite{cao2019multi} for driving fatigue, and Dataset III \cite{data4010014} for cognitive workload.

\noindent\textbf{Dataset I}: Dataset I provides EEG signals recorded while subjects perform the Discrimination/Selection Response (DSR) task to assess cognitive attention. The experiment involves a total of 26 participants. Each subject undergoes three sessions, with each session consisting of multiple cycles of attention tasks, each lasting 40 seconds, and alternating with rest periods of 20 seconds. Data collection includes recordings from 28 EEG channels, sampled at a rate of 1000 Hz. BrainAmp EEG amplifier (Brain Products GmbH, Gilching, Germany) was used to collect the EEG signals. For the pre-processing, a band-pass filter ranging from 0.5 to 50 Hz was applied to eliminate low and high-frequency noise, following the methodology in \cite{9361688}. Eye movement artifacts were removed using the automatic Independent Component Analysis (ICA) Electrooculography (EOG) removal method in the MNE toolbox \cite{GramfortEtAl2014}. The data were then downsampled to 200 Hz. In accordance with \cite{10025569}, only the first half of each attention trial (20 seconds) was used to balance the samples between attention and inattention (rest). Each trial was further segmented into 4-second segments with a 50\% overlap. All three sessions are utilized in this study. 

\noindent\textbf{Dataset II}: Dataset II is designed to evaluate cognitive fatigue states in drivers through EEG signal analysis during a 90-minute driving task within a virtual reality (VR) driving environment. This dataset was compiled with the participation of 27 subjects. EEG data were captured using a 32-channel Scan SynAmps2 Express system (Compumedics Ltd., VIC, Australia), with a sampling rate of 500 Hz. The official pre-processed data are used. For pre-processing, the raw EEG signals were filtered by a band-pass filter from 1 to 50 Hz. Eye blinks were manually removed through visual inspection. Ocular and muscular artifacts were removed using the Automatic Artifact Removal (AAR) method in EEGLab \cite{DELORME20049}. Following \cite{10025569}, the data were downsampled to 128 Hz. For calculating fatigue levels, we also adhered to the approach outlined in \cite{10025569}. EEG trials were defined as the 3 seconds of EEG data preceding the onset of lane-departure events. Reaction time (RT) was used to measure fatigue levels for these EEG trials. RT was defined as the interval from the onset of the lane-departure event to the onset of the counter-steering event. The local RT ($RT_{l}$) of a trial was defined as the RT for that specific trial. The global RT ($RT_{g}$) of a trial was the mean of the local RTs of all trials within a 90-second window before the current trial. The 5th percentile of all local RTs in the entire session was selected as the alert RT ($RT_{a}$). The labeling process was defined as follows:
\begin{equation}\label{eq:fatigue-labeling}
     y = \left\{\begin{matrix}
0 & RT_{l} > 2.5*RT_{a} \&\& RT_g > 2.5*RT_{a}\\ 
1 & RT_{l} < 1.5*RT_{a} \&\& RT_g < 1.5*RT_{a}
\end{matrix}\right.
 \end{equation}
where 0 represent the fatigue class and 1 represent the non-fatigue class. Following \cite{10025569}, 11 subjects who had enough samples for each class are utilized. 

\noindent\textbf{Dataset III}: Dataset III provides 19-channel EEG recordings from 36 subjects engaged in mental cognitive tasks, specifically performing serial subtraction, along with their corresponding baseline EEG for reference. The official artifact-free data are used in this study. The EEG were recorded using a Neurocom EEG 23-channel system (Ukraine, XAI-MEDICA). Electrodes were positioned on the scalp according to the International 10-20 system, and all electrodes were referenced to ear reference electrodes. A high-pass filter with a 30 Hz cut-off frequency and a power line notch filter (50 Hz) were applied.  ICA was used to eliminate artifacts. Each mental workload trial lasts for 60 seconds, and the last 60 seconds of the rest EEG are used as the low workload data. The trials are segmented using a 4-second sliding window with a moving step of 2 seconds. 

\begin{table*}[htp] \centering\arraybackslash
\caption{Effects of different information purification methods on three benchmarking datasets, with the best results highlighted in bold.}
\label{tab:ip_results}
\begin{adjustbox}{center}
\begin{tabular}
{lwc{2.8em}wc{2.8em}wc{2.8em}wc{2.8em}wc{2.8em}wc{2.8em}wc{2.8em}wc{2.8em}wc{2.8em}wc{2.8em}wc{2.8em}wc{2.8em}}

% {lwc{2.3em}wc{2.3em}wc{2.3em}wc{2.3em}wc{2.3em}wc{2.3em}wc{2.3em}wc{2.3em}wc{2.3em}wc{2.3em}wc{2.3em}wc{2.3em}}
\toprule
\multirow{2}{*}{Method} & \multicolumn{4}{c}{Dataset I: Attention} & \multicolumn{4}{c}{Dataset II: Fatigue} & \multicolumn{4}{c}{Dataset III: Mental Workload} \\ \cmidrule(lr){2-5} \cmidrule(lr){6-9} \cmidrule(lr){10-13}
                & ACC & std & F1-macro& std& ACC& std & F1-macro & std & ACC& std & F1-macro & std\\ \midrule
		     Using mean & 77.35 & 6.99 & 76.99 & 7.33 & 73.82 & 12.13 & 71.72 & 12.93 & 68.82 & 15.44 & 65.47 & 18.90 \\
              Using std & 77.54 & 7.54 & 77.04 & 8.07 & 73.94 & 14.42 & 72.34 &  14.77 & 72.27 & 15.20 & 69.19 & 19.00\\
              Using power &\textbf{82.72} &8.00 &\textbf{82.36}	&8.52	&\textbf{79.32} &7.87 &\textbf{75.83}	&11.54 &\textbf{73.18} &15.63	&\textbf{69.99}	&19.80\\ \bottomrule
			\end{tabular}
		\end{adjustbox}
	\end{table*}

 \begin{table*}[htp] \centering\arraybackslash
\caption{Effects of various information purification locations on three benchmarking datasets, with the best results highlighted in bold.}
\label{tab:ip_location_results}
\begin{adjustbox}{center}
\begin{tabular}
{lwc{2.8em}wc{2.8em}wc{2.8em}wc{2.8em}wc{2.8em}wc{2.8em}wc{2.8em}wc{2.8em}wc{2.8em}wc{2.8em}wc{2.8em}wc{2.8em}}

% {lwc{2.3em}wc{2.3em}wc{2.3em}wc{2.3em}wc{2.3em}wc{2.3em}wc{2.3em}wc{2.3em}wc{2.3em}wc{2.3em}wc{2.3em}wc{2.3em}}
\toprule
\multirow{2}{*}{Method} & \multicolumn{4}{c}{Dataset I: Attention} & \multicolumn{4}{c}{Dataset II: Fatigue} & \multicolumn{4}{c}{Dataset III: Mental Workload} \\ \cmidrule(lr){2-5} \cmidrule(lr){6-9} \cmidrule(lr){10-13}
                & ACC & std & F1-macro& std& ACC& std & F1-macro & std & ACC& std & F1-macro & std\\ \midrule
		     At $\boldsymbol{F}_{i+1}$ & 80.28 & 7.46 & 79.81 & 8.03 &  77.25 &  7.14 & 73.98 &  9.96 & 69.16 & 15.46 & 66.35 & 18.19\\
              At  $\boldsymbol{F}_{cg}^{i}$& 72.90 & 7.30 & 72.05 & 8.35 & 44.72 & 16.48 & 29.96 & 8.30 & 71.41 & 15.30 & 68.56 & 18.50\\
              At $\boldsymbol{F}_{fg}^{i}$ &\textbf{82.72} &8.00 &\textbf{82.36}	&8.52	&\textbf{79.32} &7.87 &\textbf{75.83}	&11.54 &\textbf{73.18} &15.63	&\textbf{69.99}	&19.80\\
               \bottomrule
			\end{tabular}
		\end{adjustbox}
	\end{table*}

\subsection{Baselines}

We demonstrate the performance of EEG-Deformer by comparing it with the following baseline methods: 
1) two graph-based methods, DGCNN \cite{8320798} and LGGNet \cite{10025569}; 2) two CNN-based methods, EEGNet \cite{Lawhern_2018} and TSception \cite{9762054}; 3) four Transformers, EEG-ViT, our adapted version of \cite{dosovitskiy2021an}, SSVEPformer \cite{CHEN2023521}, EEG-Transformer \cite{lee2022eeg}, and EEG-Conformer \cite{9991178}.

\subsubsection{DGCNN} 
DGCNN introduces a learnable adjacency matrix that dynamically captures and updates relationships between EEG channels during training, allowing for a more flexible and accurate representation of the EEG signal.
\subsubsection{LGGNet}
LGGNet, a neurology-inspired graph neural network, captures local-global-graph representations of EEG data. Its input layer uses multi-scale temporal convolutions with attentive fusion to capture EEG dynamics, followed by graph-filtering layers that model complex relationships within and between brain regions.
\subsubsection{EEGNet}
EEGNet is a compact convolutional neural network designed for EEG-based BCIs, using depth-wise and separable convolutions to efficiently integrate EEG feature extraction techniques, making it highly effective for EEG decoding.
\subsubsection{TSception}
TSception is a multi-scale convolutional neural network that combines dynamic temporal layers, asymmetric spatial layers, and advanced fusion to capture complex patterns in EEG signals. This design effectively extracts discriminative features for accurate emotion recognition.
\subsubsection{EEG-ViT}
To explore the effects of combining CNNs and Transformers for decoding EEG signals, we adapt ViT \cite{dosovitskiy2021an}, a purely Transformer-based architecture, for EEG data. We partition the EEG into non-overlapping segments along the temporal dimension and convert these segments into tokens using a linear projection layer, enabling ViT to process EEG effectively.
\subsubsection{SSVEPformer} 
The SSVEPformer is a Transformer-based method that comprises channel combination, SSVEPformer encoder, multilayer perceptron (MLP) head three core components that can extract spectrum information from the complex spectrum representation of EEG. 
\subsubsection{EEG-Transformer}
EEG-Transformer is a convolutional Transformer that combines an EEGNet-based CNN feature extractor with a vanilla Transformer \cite{NIPS2017_3f5ee243} to decode EEG signals effectively.

\subsubsection{EEG-Conformer}
EEG-Conformer, like EEG-Transformer, combines CNNs and Transformers to capture spatial-temporal information in EEG signals. It uses 1-D CNNs for local feature extraction and a self-attention module to reveal global correlations in these temporal features.

\subsection{Experiment Settings}
In this study, we implement generalized subject-independent settings, ensuring that test data information is not used during the training stage. We adopt a leave-one-subject-out (LOSO) approach for all three datasets. In each LOSO step, one subject's data is set aside as test data. Of the remaining data, 80\% is used for training and the remaining 20\% serves as validation (development) data. We employ Accuracy (ACC) and Macro-F1 as our evaluation metrics which can be formularized as
\begin{equation}\label{eq:acc}
     Accuracy = \frac{TP+TN}{TP+FP+TN+FN}
 \end{equation}

 \begin{equation}
      F1\text{-macro} = \frac{1}{N} \sum_{i=1}^{N} F1\text{-score}_i,
 \end{equation}

\begin{equation}
      F1\text{-score}_i = \frac{2 \cdot TP_i}{2 \cdot TP_i + FP_i + FN_i},
\end{equation}
where $TP$ is the true positive, $TN$ is the true negative, $FP$ is the false positive, $FN$ is the false negative, and the subscript indicates the $i$-th class.

\begin{figure*}[t]
\centering
    \subfigure[]{
    \includegraphics[width=0.13\linewidth]{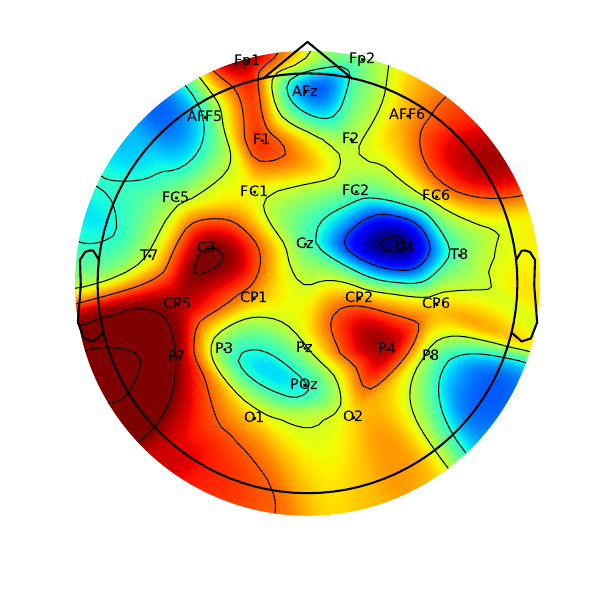}
    }
    \subfigure[]{
    \includegraphics[width=0.13\linewidth]{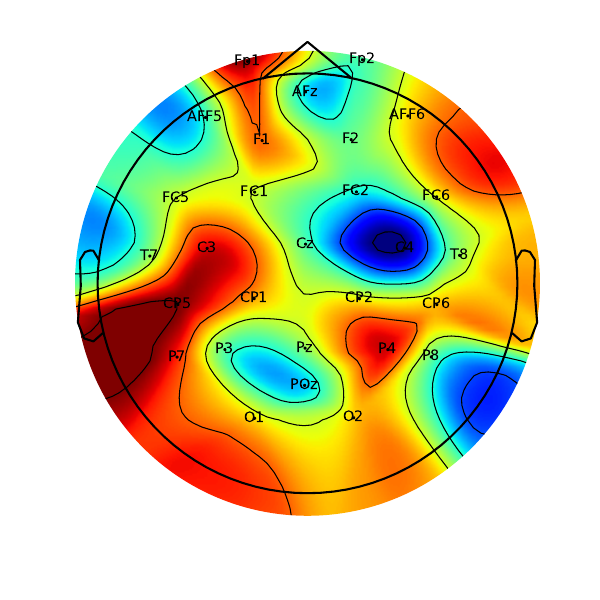}
    }
    \subfigure[]{
    \includegraphics[width=0.13\linewidth]{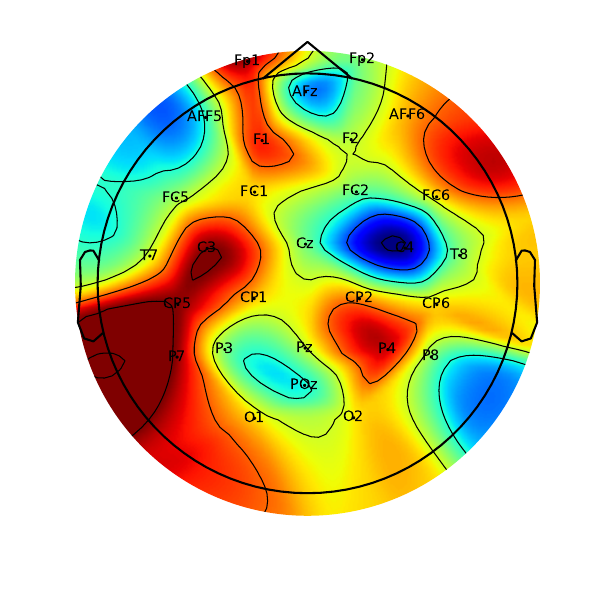}
    }
    \subfigure[]{
    \includegraphics[width=0.13\linewidth]{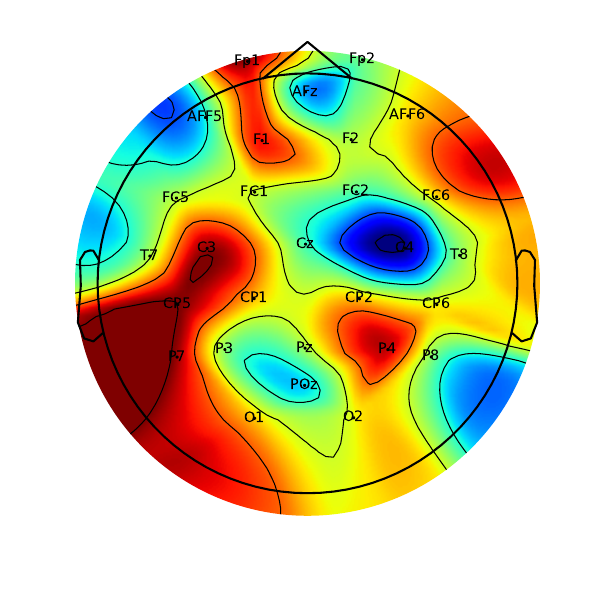}
    }
    \subfigure[]{
    \includegraphics[width=0.13\linewidth]{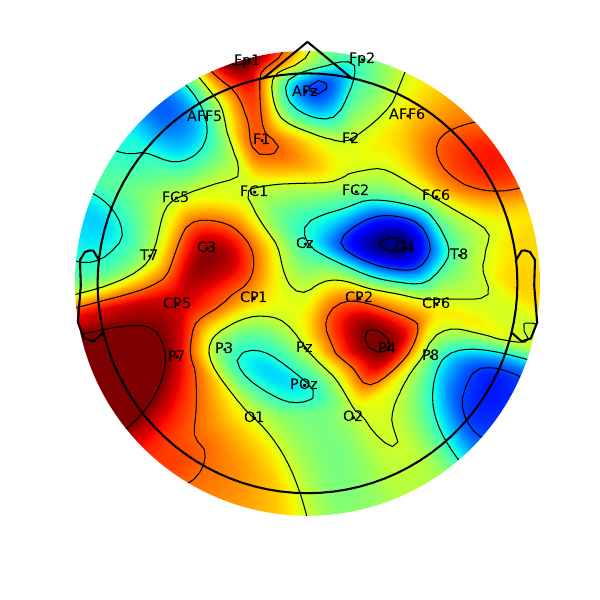}
    }
    \subfigure[]{
    \includegraphics[width=0.13\linewidth]{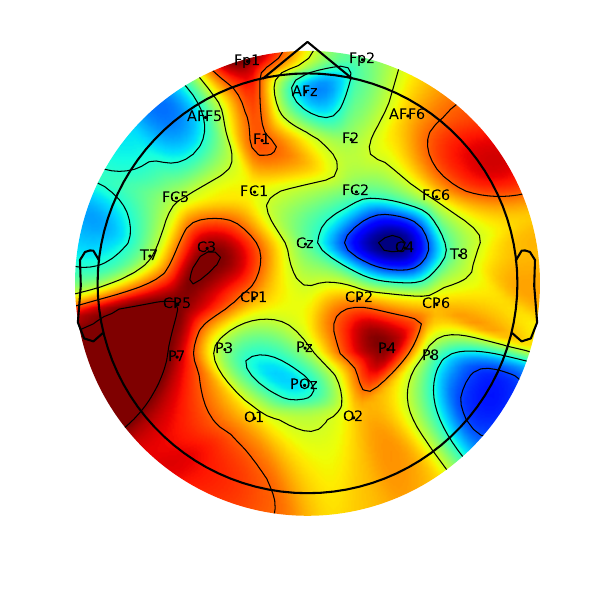}
    }
\caption{Saliency maps of attention classification tasks. (a)-(e) are five representative subjects and (f) is the average of all the subjects. The most informative areas are primarily located in the frontal (Fp1, F1, and AFF6) and parietal (CP5, P7, and P4) regions. The location of each EEG electrode can be found according to its name on the saliency map.}
\label{fig:smap_attention}
\end{figure*}

\begin{figure*}[t]
\centering
    \subfigure[]{
    \includegraphics[width=0.13\linewidth]{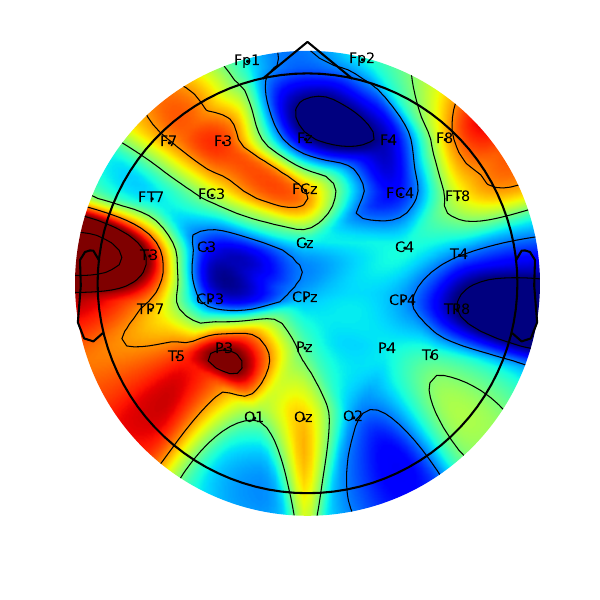}
    }
    \subfigure[]{
    \includegraphics[width=0.13\linewidth]{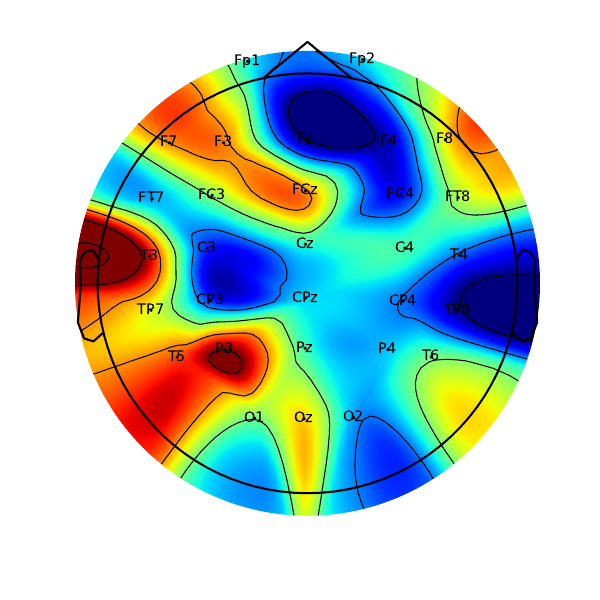}
    }
    \subfigure[]{
    \includegraphics[width=0.13\linewidth]{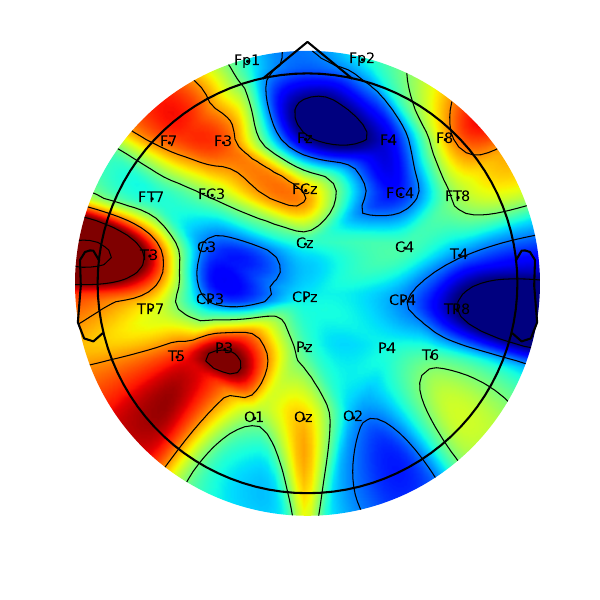}
    }
    \subfigure[]{
    \includegraphics[width=0.13\linewidth]{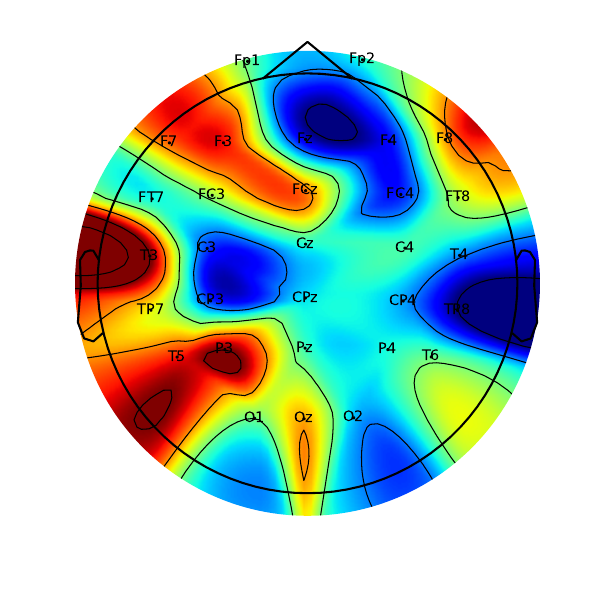}
    }
    \subfigure[]{
    \includegraphics[width=0.13\linewidth]{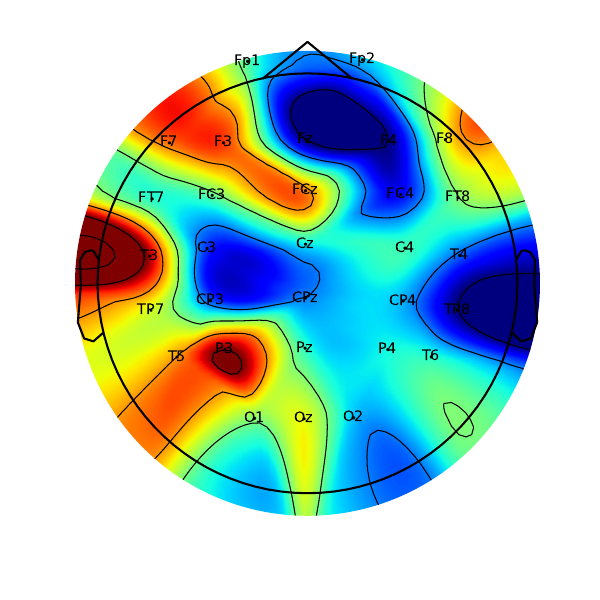}
    }
    \subfigure[]{
    \includegraphics[width=0.13\linewidth]{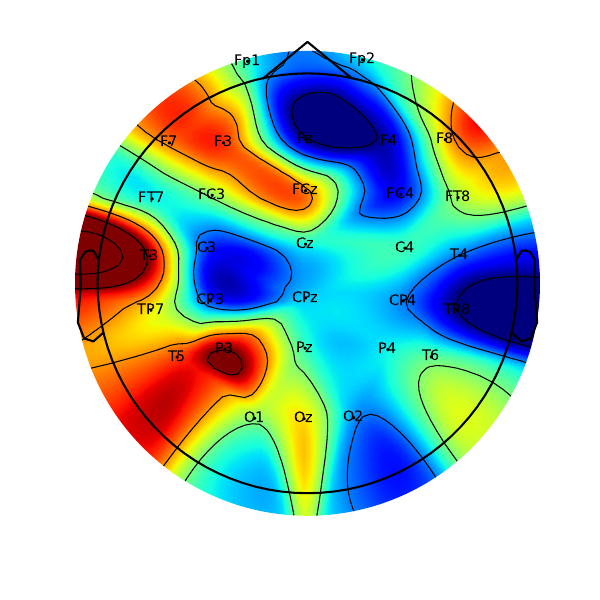}
    }
\caption{Saliency maps of fatigue classification tasks are shown. Figures (a) to (e) represent five representative subjects, and figure (f) is the average of all the subjects. The most informative areas are the frontal (F7, F3, and FCz), temporal (T3 and T5), and parietal (P3) regions. The location of each EEG electrode can be identified by its name on the saliency map.}
\label{fig:smap_fatigue}
\end{figure*}

\begin{figure*}[t]
\centering
    \subfigure[]{
    \includegraphics[width=0.13\linewidth]{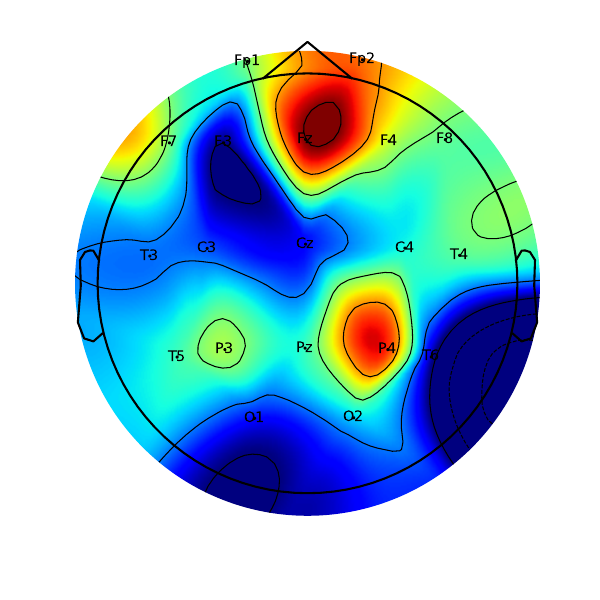}
    }
    \subfigure[]{
    \includegraphics[width=0.13\linewidth]{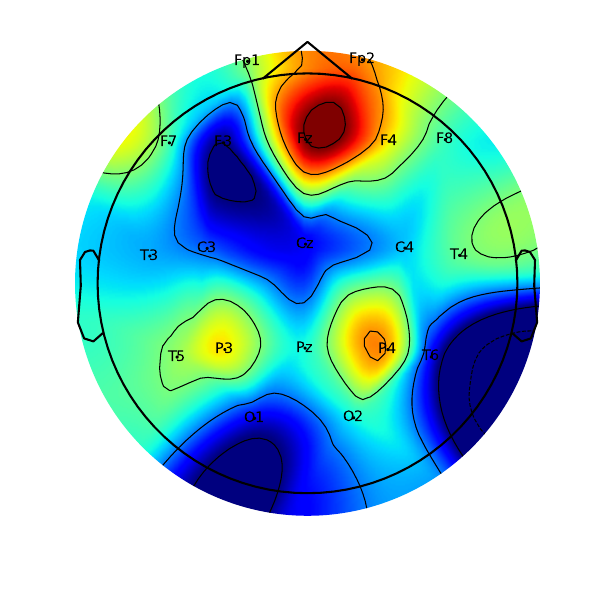}
    }
    \subfigure[]{
    \includegraphics[width=0.13\linewidth]{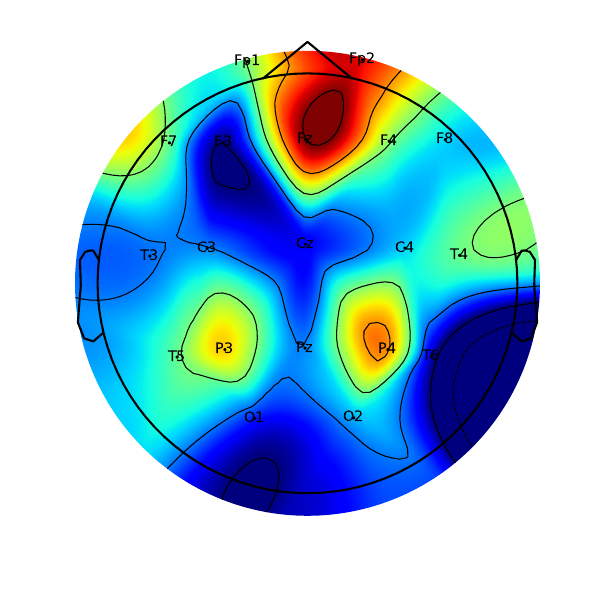}
    }
    \subfigure[]{
    \includegraphics[width=0.13\linewidth]{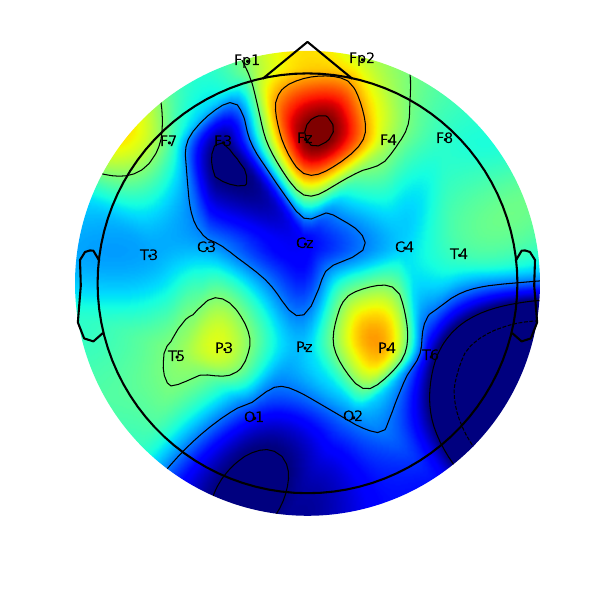}
    }
    \subfigure[]{
    \includegraphics[width=0.13\linewidth]{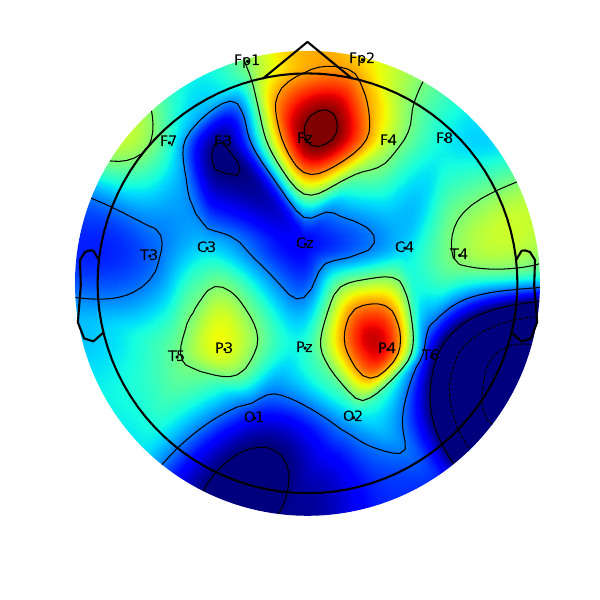}
    }
    \subfigure[]{
    \includegraphics[width=0.13\linewidth]{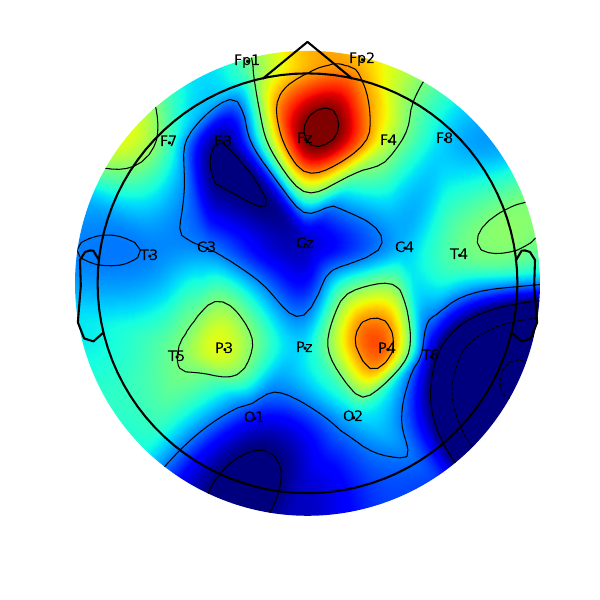}
    }
\caption{Saliency maps for mental workload classification tasks are presented. Figures (a) to (e) represent five representative subjects, and figure (f) shows the average of all subjects. The frontal (Fz and Fp2) and parietal (P4) areas are found to be more informative. The location of each EEG electrode can be identified by its name on the saliency map.}
\label{fig:smap_mwl}
\end{figure*}

\begin{figure}[htp]
\centering
    \subfigure[Dropped ACC]{
    \includegraphics[width=\linewidth]{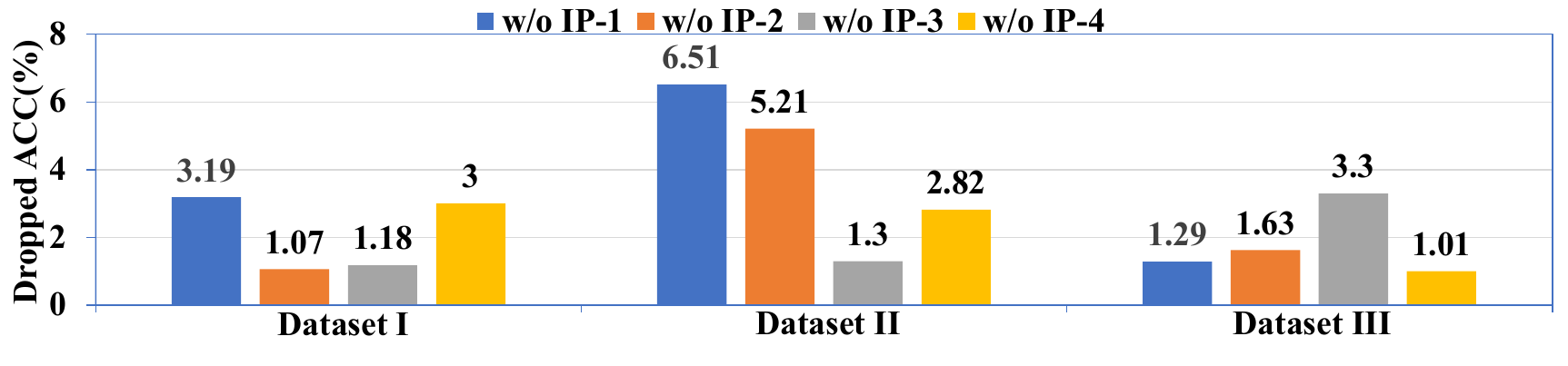}
    }
    \subfigure[Dropped F1]{
    \includegraphics[width=\linewidth]{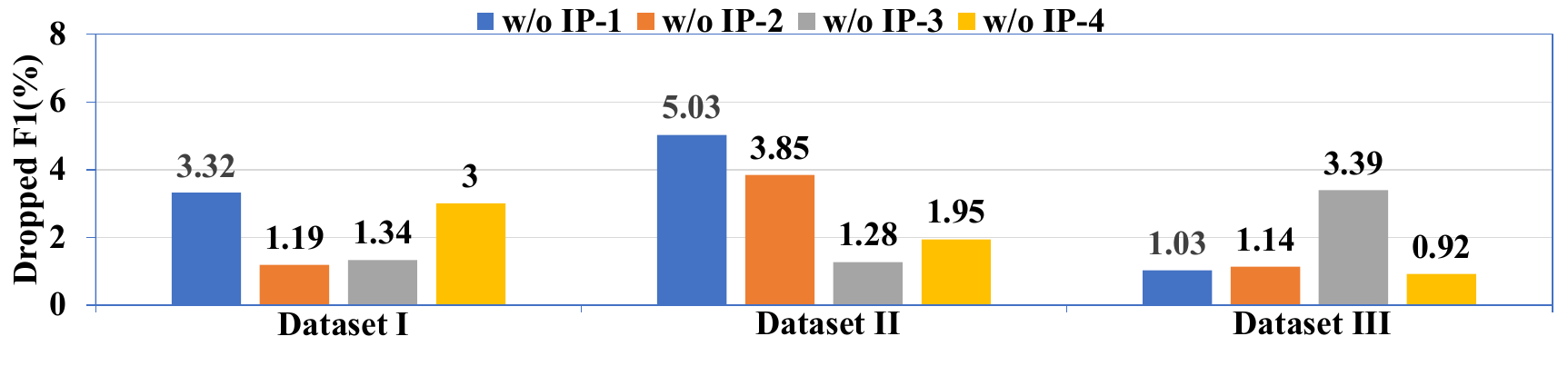}
    }

\caption{Dropped ACC (a) and F1 scores (b) resulted from removing each IP-Unit from EEG-Deformer. Removing the first IP-Unit significantly impacted performance on datasets I and II, while the third IP-Unit was vital for dataset III, reflecting their input EEG segment lengths of 800, 384, and 2000, respectively. This indicates that longer EEG inputs require deeper IP-Units.}
\label{fig:ablation_IP}
\end{figure}

\subsection{Implementation Details}

The PyTorch implementation is available on this website\footnote{https://github.com/yi-ding-cs/EEG-Deformer}. The cross-entropy loss is utilized to guide the training. We use an Adam optimizer with an initial learning rate of 1e-3 and a weight decent of 1e-5. A cosine annealing schedule is adopted to adjust the learning rate during training. A dropout rate of 0.5 is applied on dataset I and II to avoid over-fitting. The dropout rate is changed to 0.25 on dataset III as it achieves a better performance on development set. The batch size is 64 for all the datasets.We train the model for 200 epochs, and the model with the best validation accuracy is used to evaluate the test data. The kernel lengths of CNN layers are calculated by $0.1*f_{s}$, where $f_{s}$ is the sampling rate of the EEG segment. The odd length is need for PyTorch to achieve the same padding. As the $f_{s}$ of three datasets are 200 Hz, 128 Hz, and 500 Hz, the kernel lengths of CNN layers $l_{kernel}^{\textrm{I}} = 21$, $l_{kernel}^{\textrm{II}} = 13$, and $l_{kernel}^{\textrm{III}} = 51$. We set the number of CNN kernels as 64 for all the datasets as \cite{10025569}. We set $n_{head} = d_{atten}$, and tune it based on the performance on the validation (development) set. Hence, we have $n_{head}^{\textrm{I}}=32$ and $n_{head}^{\textrm{II}} = n_{head}^{\textrm{III}} = 16$.

\section{Results and Analyses}
\subsection{Classification Results}

\noindent\textbf{Attention Classification}: The results for attention classification using dataset I, as presented in Table \ref{tab:classification_results}, reveal that the EEG-Deformer outperforms all baseline methods. It achieves the highest accuracy and macro-F1 score in attention classification with 82.72\% accuracy and a 82.36\% macro-F1 score. Compared to EEG-Conformer, the next best performer, EEG-Deformer shows an improvement of 2.92\% in accuracy and 3.35\% in macro-F1 score. The results highlight that CNN-based methods generally outperform GNN-based ones and suggest that a combination of CNN and Transformer architectures yields better performance than using either CNN or Transformer alone. 

\noindent\textbf{Fatigue Classification}: The same observation appears in the fatigue decoding task using dataset II, as detailed in Table \ref{tab:classification_results}, the EEG-Deformer demonstrates superior performance, leading with an accuracy of 79.32\% and a macro-F1 score of 75.83\%. These results surpass those of the EEG-Conformer by 4.96\% in accuracy and 4.18\% in macro-F1 score. The same trend is shown here that the CNN-based methods are better than GNNs. And using Transformer without CNN layers yield low classification results. 

\noindent\textbf{Mental Workload Classification}: Regarding the classification of mental workload using Dataset III, the EEG-Deformer continues to outperform its counterparts, achieving an accuracy of 73.18\% and a macro-F1 score of 69.99\%. Unlike in other tasks, the CNN-based method TSception ranks second here, closely followed by EEG-Conformer. 

As a summary: 1) RNN-based methods is inferior to CNN-Transformers, EEG-Conformer, as indicated in \cite{9991178}. Because EEG-Conformer has a better hierarchical sequence learning ability. 2) GNN-based methods have worse performance as they depend heavily on spatial connectivity patterns, which can vary across subjects \cite{tompson2018network}, and are further constrained by EEG's high temporal but low spatial resolution. 3) The results show that Convolutional Transformers excel over traditional Transformers in EEG classification tasks, which is consistent with the observations in \cite{9845479}. 4) Although the EEG-Conformer outpaces other baselines, our model surpasses it by effectively capturing coarse-to-fine temporal information and utilizing multi-layer insights. Our enhancements specifically address these gaps, advancing EEG data analysis by learning coarse-to-fine temporal information and utilizing purified information from each Transformer layer. 

% In summary, the results show that Convolutional Transformers excel over traditional Transformers in EEG classification tasks, which is consistent with the observations in \cite{9845479}. Additionally, within Convolutional Transformers, EEG-Conformer and EEG-Deformer, learning coarse-to-fine temporal information and utilizing purified information from each Transformer layer enhances classification results.

% In summary, CNN-based methods and Convolutional Transformers generally perform better than GNN-based methods. This might be because the connectives among different electrodes vary across subjects, and GNN-based methods highly rely on these connectivity patterns. Convolutional Transformers are more effective than traditional Transformers for EEG classification tasks. This might because CNNs serve as digital filters, which can capture signals across different frequency bands of EEG. Accoding to the comparison within convolutional Transformers, effectively learning the coarse-to-fine temporal information in Transformer layers and using the purified information from each Transformer layer can improve the classification results. 

\subsection{Ablation}\label{sct:ablation}

To investigate the individual contributions of the FTL, Dense connections of IP-Units, and IP-Unit, we conduct an ablation study by selectively removing these layers and observing their impact on classification outcomes across three datasets. The findings, detailed in Table \ref{tab:ablation_results}, indicate that the omission of the FTL had the most detrimental effect on performance overall. Additionally, the removal of the IP-Unit results in the second-largest decrease in performance on datasets I and II. Eliminating dense connections also leads to reductions in both classification accuracy and macro-F1 scores across all datasets. These results collectively imply that each module within the EEG-Deformer plays a crucial role, working in unison to enhance its predictive capabilities.

\begin{figure}[htp]
\centering
    \subfigure[Dataset I]{
    \includegraphics[width=0.82\linewidth]{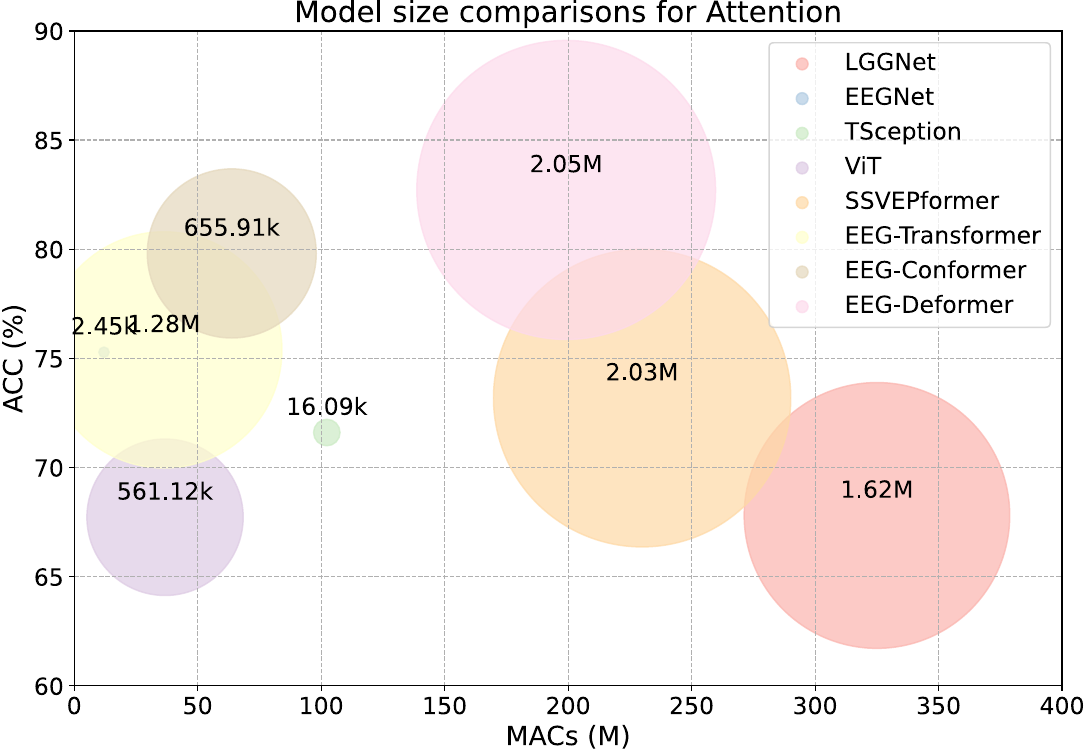}
    }
    \subfigure[Dataset II]{
    \includegraphics[width=0.82\linewidth]{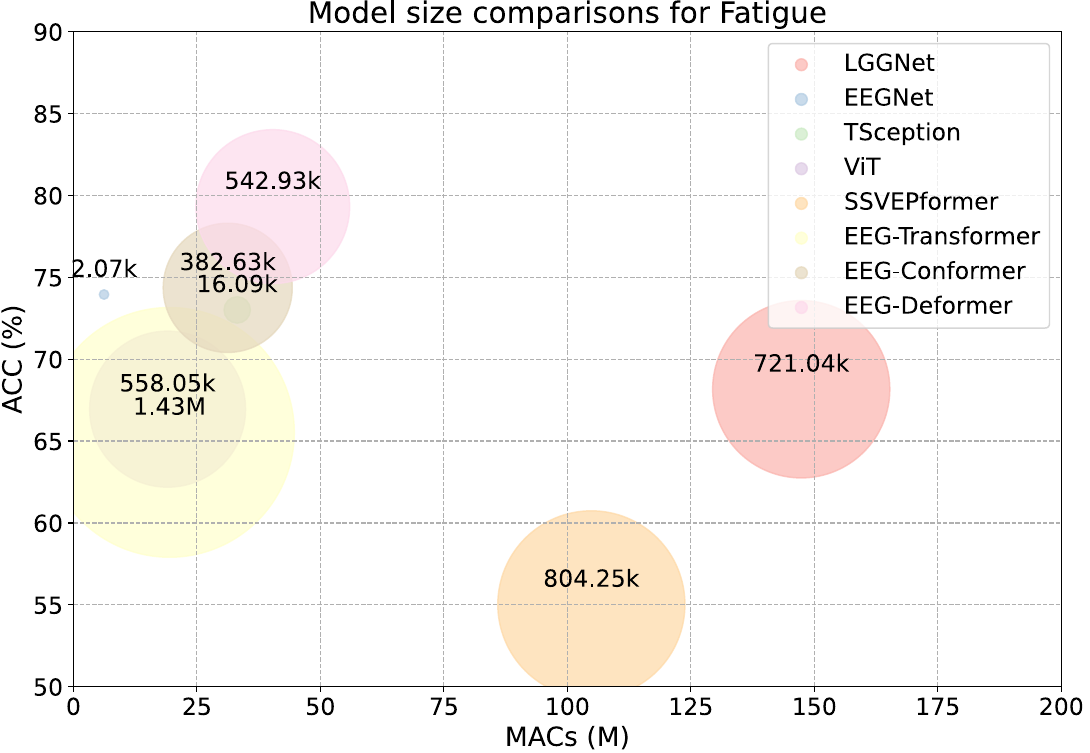}
    }
     \subfigure[Dataset III]{
    \includegraphics[width=0.82\linewidth]{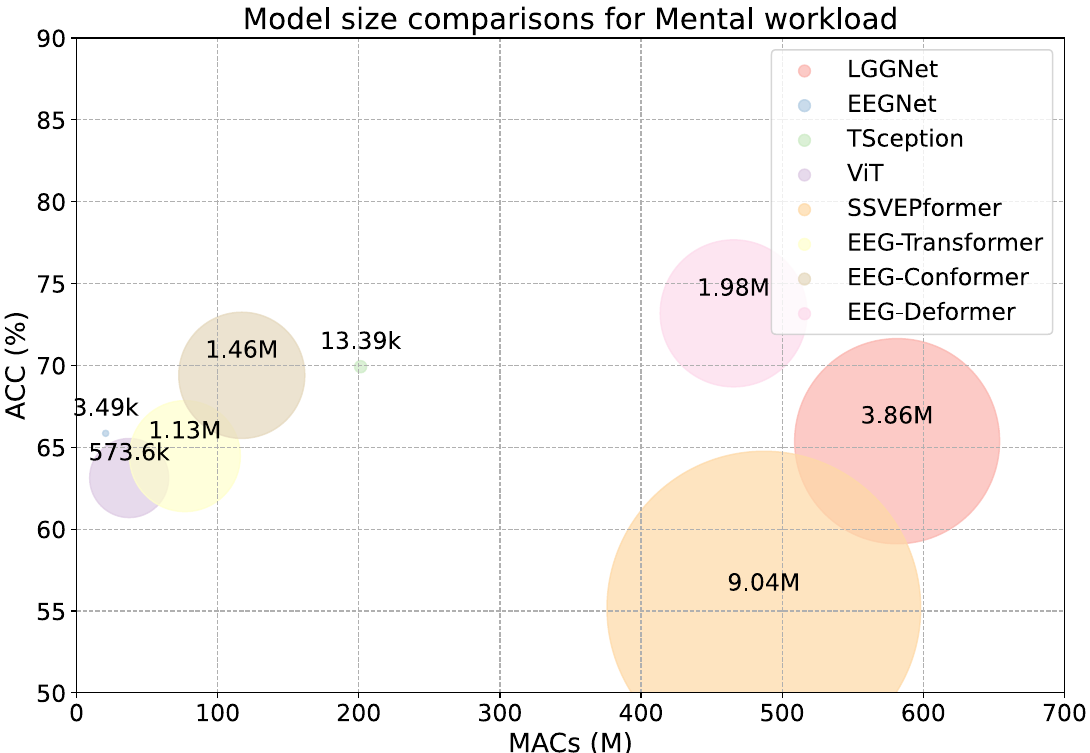}
    }

\caption{Comparisons about performance (ACC, on y-axis), model size (size of bubbles), and computational complexity (MACs, on x-axis). Light pink one is Deformer (ours). Our EEG-Deformer strikes a promising balance between accuracy and computational complexity.}
\label{fig:model_size}
\end{figure}

\subsection{Ablation of IP-Units}\label{sct:IP_types}
To further understand the role of each IP-Unit, we conducted a removal analysis where each IP-Unit was individually omitted to observe its impact. Figure \ref{fig:ablation_IP} depicts the consequences of this removal, displaying reductions in accuracy (ACC) and F1 score. The results show that eliminating the first IP-Unit resulted in the largest declines on datasets I and II, whereas the third IP-Unit was more crucial for dataset III. These findings seem to correlate with the lengths of the input EEG segments, which are 800, 384, and 2000 for datasets I, II, and III, respectively. This suggests that the importance of deeper IP-Units increases with the length of the input EEG segment. Additionally, removing the last IP-Unit also led to significant reductions in both ACC and macro-F1 on datasets I and II, underscoring that all IP-Units collectively contribute to performance enhancements.

\subsection{Effect of Different Types of IP-Unit}\label{sct:IP_types}

Table \ref{tab:ip_results} illustrates the impact of different information purification methods, which are mean, standard deviation (std), and power on the performance across three datasets. For each layer's output, $\textbf{\textit{F}}_{fg}^{i} \in \mathbb{R}^{k \times 0.5*l_{t}^{i}}$, we will calculate the mean, std on the last dimension of $
\textbf{\textit{F}}_{fg}^{i}$, resulting the purified information, $\textbf{\textit{I}}_{i} \in \mathbb{R}^{k}$. The comparison focuses on their effects on accuracy and macro-F1 score. Notably, the 'power' method consistently yields the highest accuracy and macro-F1 scores across all datasets, outperforming the other two methods. These findings underscore the superiority of using the 'power' method for information purification in this context.

\subsection{Effect of Different Locations of The IP-Unit}\label{sct:location_IP}

The study assesses the impact of varying the placement of IP-Unit within each HCT block, focusing on three specific locations in Figure~\ref{fig:hct} : at the fused output, $\boldsymbol{F}_{i+1}$; at the coarse-grained representation, $\boldsymbol{F}_{cg}^{i}$; and at the fine-grained representation, $\boldsymbol{F}_{fg}^{i}$. The results of this evaluation are presented in Table~\ref{tab:ip_location_results}. These findings reveal that applying IP-Unit to the output of the FTL leads to the highest decoding performance across all three datasets. This outcome highlights the effectiveness of this particular placement for IP-Unit within the HCT block structure.

\subsection{The Parameters Comparison of EEG-Deformer and Other Baseline Methods.}
We visualize the comparison between different methods with respect to parameters, Multiply-Accumulate Operations (MACs), and accuracy. The results are shown in Fig.~\ref{fig:model_size}. We only compare models that use EEG as input, as the hand-crafted features have fewer data points in each input. On dataset I, the EEG-Deformer requires fewer MACs than models of similar size, while achieving higher accuracy than smaller models. On dataset II, the EEG-Deformer has the highest performance with a relatively smaller model size and fewer MACs compared to other baselines. On dataset III, although the EEG-Deformer uses more MACs, it has a smaller model size than those requiring more MACs. Generally, our EEG-Deformer achieves a promising balance between accuracy and computational efficiency.

\subsection{Visualization}
In this study, we use saliency maps \cite{simonyan2013deep} to visualize the regions that the neural networks identify as most informative. Figures \ref{fig:smap_attention} to \ref{fig:smap_mwl} show saliency maps for five randomly selected subjects, as well as the averaged map across all subjects, for three datasets corresponding to attention, fatigue, and mental workload classification tasks. To improve clarity, these saliency maps are normalized to a range between 0 and 1. The brain is functionally divided into frontal, temporal, parietal, and occipital regions. According to Figure \ref{fig:smap_attention}, the areas most informative for attention classification are primarily located in the frontal (Fp1, F1, and AFF6) and parietal (CP5, P7, and P4) regions, which are known to be related to cognitive attention \cite{10.1093/cercor/bhu204}. Figure \ref{fig:smap_fatigue} shows that for fatigue classification, the neural network primarily focuses on the frontal (F7, F3, and FCz), temporal (T3 and T5), and parietal (P3) areas, consistent with \cite{FLORHENRY2010155}. In the case of mental workload classification, we find consistent findings with those reported in \cite{9444793}: the frontal (Fz and Fp2) and parietal (P4) areas are more informative for neural networks, as shown in Figure \ref{fig:smap_mwl}.
These visualizations align with known brain activity patterns related to these cognitive processes and demonstrate the neural network's ability to identify relevant brain regions for each task. As shown in the figures, the saliency maps display generally consistent patterns for each task, indicating that our model consistently learns from task-related brain regions.

\subsection{Limitations and future work}
While the model demonstrates strong performance on the selected tasks, the evaluation does not cover all commonly used tasks in EEG-based classification. Expanding the range of tasks could provide a more comprehensive assessment of the model’s generalizability across different applications. In some datasets, the standard deviation of the model’s performance is higher compared to certain baseline methods. This suggests that, although the model performs well on average, it may exhibit greater variability across different subjects or trials, potentially affecting its reliability in certain cases. In the future, efforts should focus on improving the model's robustness.

\section{Conclusion}

In this paper, we introduce EEG-Deformer, a novel convolutional Transformer designed for cross-subject EEG classification tasks. The model begins with a shallow CNN encoder, followed by a series of HCT blocks. These blocks are specifically designed to capture both coarse- and fine-grained temporal dynamics from EEG signals, achieved by integrating a FTL into the Transformers. To leverage multi-level temporal information effectively, the model performs information purification on the fine-grained temporal representations extracted by all HCT layers. These purified signals are then densely connected to the final embeddings.
We evaluate EEG-Deformer against various baselines across three benchmark datasets. The results consistently demonstrate that EEG-Deformer outperforms these baselines, showcasing its effectiveness in EEG decoding tasks. These findings suggest that EEG-Deformer can serve as a robust and versatile backbone for a wide range of EEG decoding tasks.  

\section*{References}

\bibliographystyle{./IEEEtran}
\bibliography{./mybib}

\end{document}